\documentclass[12pt]{article}
\usepackage{color}
\usepackage{graphics}
\usepackage{amsfonts}
\usepackage{epsfig}
\def\one{{\hbox{1\kern-.8mm l}}}

\def\t{\tau }

\newcommand{\beq}{\begin{equation}}
\newcommand{\eeq}{\end{equation}}
\newcommand{\be}{\begin{eqnarray}}
 \newcommand{\ee}{\end{eqnarray}}

 \newcommand{\half}{\frac{1}{2}}

\newcommand{\ov } {\over }
\newcommand{\p }{\partial }

\newcommand{\s }{\sigma }
 
\newcommand{\T }{ {\cal T } }

\def\td{\tilde }
\def\a{\alpha }

\begin{document}
%%%%%%%%%%%%%%%%%%%%%%%%%%%%%%%%
\null\vskip-24pt 
\hfill
SISSA-94/2003/EP 
\vskip-1pt
\hfill {\tt hep-th/0310283}
\vskip0.2truecm
\begin{center}
\vskip 0.2truecm {\Large\bf
Decay of long-lived massive closed\\
\vskip 0.2truecm
 superstring states: Exact results}
\\
\vskip 2.5truecm

{\bf Diego Chialva $^a$
, Roberto Iengo $^{a,b}$ 
and Jorge G. Russo $^{c}$
}

\bigskip
\medskip

%\centerline{${}^a$

{{${}^a$
\it International School for Advanced Studies (SISSA)\\
Via Beirut 2-4, I-34013 Trieste, Italy} 

%\centerline
\medskip

{${}^b$ INFN, Sezione di Trieste}

%\centerline

\medskip

{${}^c$  
Instituci\' o Catalana de Recerca i Estudis Avan\c{c}ats (ICREA),\\
Departament ECM,
Facultat de F\'\i sica, Universitat de Barcelona,  Spain}
}
\bigskip

{\tt{chialva@sissa.it},
{iengo@he.sissa.it},
{jrusso@ecm.ub.es}
}

\end{center}
\begin{abstract}

We find a one-parameter family of long-lived physical
string states in type II superstring theory.
We compute
the decay rate by an exact numerical evaluation of the
imaginary part of the one-loop propagator. Remarkably,  
the lifetime rapidly increases with the mass. We find a
power-law dependence of the form 
$\T \cong {\rm const.} g_s^{-2}\ ({\rm mass })^\alpha$,
where the value of $\alpha $ depends on the parameter
characterizing the state. For the most stable state in this family, 
one has $\alpha\cong 5$. 
The dominant decay channel of these massive string states is by 
emission of soft massless particles.
The quantum states  can be viewed semiclassically as 
closed strings which cannot break during the classical evolution.

% $\alpha \in (\alpha_0,\alpha_1)$, $\alpha_0\cong  0.1$,
% $\alpha_1\cong 4.9$.

% The states can be interpreted semiclassically as rotating ellipses.
%We argue that there ma be many other long-lived states in
%the theory.

\end{abstract}

\date{October 2003}

\vfill\eject

\tableofcontents
\listoftables
\listoffigures

\setcounter{section}{0}

%\begin{document}
%\maketitle

%%%%%%%%%%%%%%%%%%%%%%%%%%%%%%%%%%%%%%%%%%%%%%%%
\section{Introduction}
%%%%%%%%%%%%%%%%%%%%%%%%%%%%%%%%%%%%%%%%%%%%%%%%

A basic feature of 
Superstring theory is the presence of an infinite tower of massive states,
with an exponential degeneracy at high mass level.
All these states are expected to be unstable, since they can
decay by emission of massive or light particles.
An important question is what is the lifetime of massive states in
superstring theory. 
This question was investigated in a number of papers in the past, 
using different methods 
\cite{Green}~--~\cite{IR1}.
%Mitchell,Dai,Okada,Sundborg,Wilkinson,Turok,Amati2,IK,Manes,IR1}. 
Nevertheless, there are no conclusive results
and the whole picture is far from
clear. 
%In particular, most studies deals only with the special 
%state of the maximum
%angular momentum, while this is only one among an exponential number of
%string states.
Could there be long-lived states in string theory?
The possible existence of quasi stable states in string theory is an
exciting subject, since it could have both theoretical as well as
phenomenological implications in Cosmology or accelerator physics in
models with large extra dimensions.

How to identify possible long-lived string states among
the vast number of states with a given mass?

In section 2 we will give a simple semiclassical argument 
which allows to find canditates for quasi stable quantum states.
The basic idea of the argument is as follows. Highly excited string
states have large quantum numbers and can be semiclassically 
described in terms of  solitons. Such solutions are classically
stable,
unless the string breaks. However, for a theory of closed string only, 
like type II superstrings, classical breaking is
possible
only if two points of the string touch during the evolution.
Thus the problem of identifying stable quantum states is mapped to the
problem
of finding classical string solutions whose points do not touch during
the evolution. Given a classical solution satisfying this requirement, 
one can write the corresponding quantum state,
and compute the lifetime explicitly by the quantum one-loop
calculation.
 When the classical breaking is not possible, 
this does not prevent that the original string can decay by emitting 
{\it light} particles (actually massless, as we will see), for which a 
semiclassical description does not apply.

With this line of argument, we have identified a one-parameter family
of  long-lived string states.

The method used here is a direct numerical evaluation of the formula
giving the imaginary part of $\Delta M^2$ at one loop.
The method is in fact similar to the one used by Okada and Tsuchiya in
\cite{Okada} for the open string state with maximum angular momentum.
In that case, it was found a lifetime $\T={\rm const.}\  M^{-1}$,
implying that the open string becomes more unstable for large masses.
We have reproduced this result as a check with our method.
However, the states we consider here have no analog
in open superstring theory, except for the less stable states of the family
which correspond to  classically breakable configurations.

In sect. 3 we explain the computation of ${\rm Im} (\Delta M^2)$,
which leads us to a formula involving a finite sum which is
to be evaluated numerically.In this paper we consider type II superstrings in 10 flat uncompactified dimensions. 

In section 4 we discuss the results of the numerical analysis.

 In the case of  those states whose classical analog is a breakable string we find that
the lifetime can increase or decrease with increasing mass, depending
on the actual quantum realization; the decay channel
into two massive pieces is dominant and the numerical results 
accurately reproduce what is expected from the classical picture of the
string breaking.

In the other cases (of quantum states which classically correspond to
unbreakable strings) the by far dominant decay channel is  the emission of
massless particles, mostly of very low energy. We find that 
the lifetime grows as a power
of the mass, and we compute the spectrum of the emitted massless particles.

In section 5 we make some additional comments, 
in particular we discuss possible gravitational effects, which were not taken in account before, 
and argue that they are not expected to play an important role.
%We also argue that gravitational effects are not expected to play an
%important role.

Some details of the calculation
are described in appendices. In appendix A, we describe the
construction of the quantum superstring states whose decay properties
are studied in this paper.
In appendix B, we construct the corresponding vertex operator. In appendix C
we compute the correlators appearing in the two-point amplitudes in
terms of theta-functions of the torus and organize the result in a
convenient way.
In appendix D we compute the fermion correlators and perform the sum
over spin structures.
In appendix E we give  a simple analytic computation
of a decay in a particular channel, obtaining
a decay rate which agrees with the numerical results.
In appendix F we show the shape of the logarithm of the decay rate divided by
$2M^2$ for different values of $M$ for some of the analysed cases.

%%%%%%%%%%%%%%%%%%%%%%%%%%%%%%%%%%%%%%%%%%%%%%%%%%%%%%%
\section{ String solitons and quantum states}

Quantum states with high quantum numbers can be described in
terms of a classical soliton. Examples on how a classical description
can describe features of the quantum decay with high accuracy were given
in ref. \cite{IR2} (see also sect. 3).
In this section we construct classical soliton solutions which
can be associated with the quantum states whose decay rates will
be computed  in the next sections. 
A classical soliton description of a quantum state can
give an intuition on the possible decay modes of the quantum state.
Here we will consider several examples and identify some
quantum states which are expected to be long-lived.

\bigskip

%%%%%%%%%%%%%%%%%%%%%%%%%%%%%%%%%%%%%%%%%%%%%%%%%%%%%%%%%
\subsection{Quantum Hilbert space}
%%%%%%%%%%%%%%%%%%%%%%%%%%%%%%%%%%%%%%%%%%%%%%%%%%%%%%%%%

We review here some standard facts of the free string
in ten flat spacetime dimensions and set the notation  
(covariant super-vertices will be discussed in the next section 
and in the appendices A,B).
Define $U=-X_0+X_9$, $V=X_0+X_9$, and set $U=2\sqrt{\a'} p_v\tau $.
Consider strings which, in addition of moving in the $X_9$ direction,
fluctuate  in the planes
$$
Z^1=\frac{X^1+i X^2}{\sqrt{2}}, \ \ Z^2=\frac{X^3+i X^4}{\sqrt{2}}\ .
$$  
We shall consider states in the NS-NS sector and omit in the whole
section 2 
the world-sheet fermions from the formulas (their inclusion is
 straightforward, but not very relevant for the discussion of this section).

The solution to the equations of motion is
\beq
V=V_L(\s^+)+V_R(\s^-)\ \ ,\ \ \ \ 
X^i(\s,\t )=X^i_L(\s^+)+X^i_R(\s^-)\ ,\ \ \ i=1,...,4\ ,
\eeq
where $\s_\pm =\s\pm \t $.
In addition, one has to solve the two constraints:
\def\p{\partial }
\be
-\sqrt{\a'} p_v \p_+ V_L &=& \p_+ Z^1_L \p_+\bar Z^1_L
 +\p_+ Z^2_L \p_+\bar Z^2_L 
\ ,\  \nonumber\\
\sqrt{\a'} p_v \p_- V_R &=& \p_-Z^1_R \p_-\bar Z^1_R
 + \p_-Z^2_R \p_-\bar Z^2_R  \ .
\label{cons}
\ee
We will work
in the center of mass frame of the $X^i$ where $p_i=0$.
{}For the complex coordinates, the Fourier mode expansions is
($\sigma\in [0,\pi )$ )
\be
Z_R^1 &=& {i} \sqrt{\frac{\a'}{2}}\sum_{n=1}^\infty {1\over \sqrt{n}}
\big[ b_{n-} e^{2in(\s -\t )} - b_{n+}^\dagger  e^{-2in(\s -\t )}\big]
\nonumber\\
Z_L^1 &=& {i} \sqrt{\frac{\a'}{2}}\sum_{n=1}^\infty {1\over \sqrt{n}}
\big[ \tilde b_{n-} e^{-2in(\s +\t )} - 
\tilde b_{n+}^\dagger  e^{2in(\s +\t )}\big]
\label{zetu}
\ee
% and a similar expansion for $Z^2_{R,L}$ with $\{b\rightarrow c\}$.
% Re: Meglio scrivere i c_n esplicitamente almeno una volta, cosi
% sono chiaramente definiti
\be
Z_R^2 &=& {i} \sqrt{\frac{\a'}{2}}\sum_{n=1}^\infty {1\over \sqrt{n}}
\big[ c_{n-} e^{2in(\s -\t )} - c_{n+}^\dagger  e^{-2in(\s -\t )}\big]
\nonumber\\
Z_L^2 &=& {i} \sqrt{\frac{\a'}{2}}\sum_{n=1}^\infty {1\over \sqrt{n}}
\big[ \tilde c_{n-} e^{-2in(\s +\t )} - 
\tilde c_{n+}^\dagger  e^{2in(\s +\t )}\big]
\label{zetdu}
\ee
The operators $b,c$ obey the usual commutation relations:
\beq
[b_{n\pm},b_{m\pm }^\dagger]=\delta_{nm}\ ,\ \ \ \ 
[\tilde b_{n\pm},\tilde b_{m\pm }^\dagger]=\delta_{nm}\ ,
\eeq
\beq
[c_{n\pm},c_{m\pm }^\dagger]=\delta_{nm}\ ,\ \ \ \ 
[\tilde c_{n\pm},\tilde c_{m\pm }^\dagger]=\delta_{nm}\ .
\eeq
The mass formula is given by
\beq
\a' M^2= 2(N_R+N_L)- a\ ,\ \ \ \ \ N_R=N_L \ ,
\label{masa}
\eeq
where the number operators are
\beq
N_R=\sum_{n=1}^\infty n( b_{n+}^\dagger b_{n+}+ b_{n-}^\dagger b_{n-}
+ c_{n+}^\dagger c_{n+}+ c_{n-}^\dagger c_{n-})\ .
\eeq
The expression for $N_L$ is similar, with the change $
\{b\rightarrow\tilde b,c\rightarrow\tilde c\}$.
%\beq
%N_L=\sum_{n=1}^\infty n( \tilde b_{n+}^\dagger \tilde b_{n+}+ 
%\tilde b_{n-}^\dagger \tilde b_{n-}+
%\tilde c_{n+}^\dagger \tilde c_{n+}+ 
%\tilde c_{n-}^\dagger \tilde c_{n-})
%\eeq
The normal ordering constant $a$ is $a=2$ in the NS-NS sector. In this
section it will be ignored since we are interested in states with
large $N_R=N_L$.

The angular momentum components 
$J_{12}=J_{12R}+J_{12L}, J_{34}=J_{34R}+J_{34L}$ 
in the $(X_1,X_2)$ and  $(X_3,X_4)$ planes
are 
\beq
J_{12R}=\sum_{n=1}^\infty ( b_{n+}^\dagger b_{n+}- b_{n-}^\dagger b_{n-})
\ ,\ \ \ \ 
J_{12L}=\sum_{n=1}^\infty ( \tilde b_{n+}^\dagger \tilde b_{n+} - 
\tilde b_{n-}^\dagger \tilde b_{n-}) \ ,
\label{espin}
\eeq
and similar expressions for $J_{34R}, J_{34L}$, 
replacing $\{ b\rightarrow c\}$.
%\beq
%S_{2R}=\sum_{n=1}^\infty ( c_{n+}^\dagger c_{n+}- c_{n-}^\dagger c_{n-})
%\ ,\ \ \ \ 
%S_{2L}=\sum_{n=1}^\infty ( \tilde c_{n+}^\dagger \tilde c_{n+} - 
%\tilde c_{n-}^\dagger \tilde c_{n-})\ .
%\label{esin}
%\eeq
The physical Hilbert space is then constructed (in the light-cone gauge)
as usual by applying the creation operators to the vacuum Fock state.
The operators $b_{n+}^\dagger,\ \tilde b_{n+}^\dagger $ raise one unit
of the spin operator $J_{12}$ in the ``up'' direction, whereas 
the operators $b_{n-}^\dagger,\ \tilde b_{n-}^\dagger $ raise one unit
of the spin operator $J_{12}$ in the ``down'' direction, and similarly
for $J_{34}$ in the $34$ plane.
\vskip1truecm

The classical string solutions have to satisfy the  constraints (\ref{cons}).
A well-known example is the rotating string with maximum angular
momentum, where the classical solution is given by
\beq
Z_1=L \cos(2\s ) e^{2i\tau }\ ,\ \ \ \ \ U=2\sqrt{\a'} p_v\tau\ ,\ \ V=2\sqrt{\a'}p_u\tau \ ,
\label{regg}
\eeq
$$
L^2=-2\a' p_u p_v={\a'}^2 M^2 \ .
$$

 The classical solution
 (\ref{regg}) is obtained by setting
\beq
-b_{1+}^{cl}  =b_{1+}^{\dagger cl } = {iL\over 2\sqrt{\a '} }\ ,
\ \ \ \ \ -\tilde b_{1+} ^{cl }=\tilde b_{1+}^{\dagger cl }=
         {iL\over 2\sqrt{\a '} }\ ,
\eeq
and 
$b_{1-}^{cl}  =b_{1-}^{\dagger cl } = 0\ ,\ \ \tilde b_{1-}^{cl} 
=\tilde b_{1-}^{\dagger cl } = 0\ .$
With these classical values for the Fourier coefficients we get 
\beq
M={L\over \a' }\ ,\ \ \ \ J={L^2\over 2\a' } \ ,
\eeq
which are the correct values of the mass $M$ and spin 
$J$ for the state of maximum angular momentum.

To have the same values of $J_R,J_L$, the corresponding quantum state  
must be of the form
\beq
|\Phi_{J_{max}} \rangle = (b_{1+}^\dagger)^N 
(\tilde b_{1+}^\dagger)^N|0\rangle \ .
\label{rott}
\eeq
This has 
$$ 
N_R=N_L=N \ .
$$ 
Since, classically, $N_R={L^2\over 4\a' }$, we have to set
$$
N={L^2\over 4\a' } \ .
$$
The soliton description applies in the large $L$ limit and
${L^2\over 4\a' }$  does not need to be an integer. In general the 
soliton is aproximately described by the  quantum state (\ref{rott}),
with $N$ being the closest integer.

%Quantum mechanically $k$ is a positive integer, and this identification
%can be done only for discrete  values of $L$. 

%%%%%%%%%%%%%%%%%%%%%%%%%%%%%%%%%%%%%%%%%%%%%%%%%%%%%%%%%%%%%%%%%%%%%%%%%%%%%
\subsection{Quantum  string states and solitons fluctuating in two planes} \label{solitonstwopl}
%%%%%%%%%%%%%%%%%%%%%%%%%%%%%%%%%%%%%%%%%%%%%%%%%%%%%%%%%%%%%%%%%%%%%%%%%%%%

In what follows we  only make use of the creation operators 
$b_{1+}^\dagger, c_{1+}^\dagger$ and 
$\tilde b_{1+}^\dagger $, $\tilde c_{1+}^\dagger $, so for clarity in
the notation we  define $b^\dagger \equiv b_{1+}^\dagger$, etc.
(i.e. we omit the subindex $\{ 1+\} $).

In this paper we shall be interested in the 
following
one-parameter family of quantum states:
% (we omit world-sheet fermions):
\beq \label{bosonsol}
|\Phi _{k,n} \rangle = {1\over (k+n)! (k-n)!} \ (b^\dagger )^{k+n}   
(c^\dagger )^{k-n}
(\tilde b^\dagger )^{k-n}   (\tilde c^\dagger )^{k+n} |0\rangle \ , \
  \ \  \langle \Phi_n |\Phi _n \rangle=1\ ,
\label{ourstate}
\eeq 
It has
\beq
N_R=N_L=2k\ ,\ \ \ \ \a ' M^2=8k\ ,\ \ \ 
\eeq
\beq
J_{12R}=k+n , \ \ \  J_{12L}=k-n  , \ \ \ \ 
J_{34R}=k-n , \ \ \  J_{34L}=k+n  , 
\eeq
so that $J_{12}=J_{34}=2k$ (the full superstring state including
world-sheet fermion modes is constructed
in appendix A).

The soliton solution with the same values of $J_{12R},J_{12L}, J_{34R},
J_{34L}$
is as follows (here $\a'=1 $)
\beq
Z_{1R}= \sqrt{k+n}\  e^{-2i(\s -\tau )}\ ,\ \ \ \ 
Z_{1L}=\sqrt{k-n}\  e^{2i(\s +\tau )} \ ,
\eeq
\beq
Z_{2R}= \sqrt{k-n}\  e^{-2i(\s -\tau )}\ ,\ \ \ \ 
Z_{2L}=\sqrt{k+n}\  e^{2i(\s +\tau )}\ ,
\eeq
or
\be
Z_1 &=& 
e^{2i\tau } \big( \sqrt{k+n}\  e^{-2i\s } + \sqrt{k-n}\  e^{2i \s }
\big)\ ,
\nonumber\\
Z_2&=&e^{2i\tau } \big( \sqrt{k-n}\  e^{-2i\s } + \sqrt{k+n}\  e^{2i \s
} \big)\ .
\label{esr}
\ee
It describes a rotating ellipse. 
It is  useful to view it in a rotated frame
\beq
Z_1'= {Z_1+Z_2 \over \sqrt{2}}= \sqrt{2}  L_1 e^{2i\tau }
 \cos(2\s )\ ,\ \ \ \ \ 
\eeq
\beq
Z_2'=  {-Z_1+Z_2 \over i\sqrt{2}}=\sqrt{2}  L_2 e^{2i\tau } \sin(2\s )\ ,
\eeq
$$
L_1= \sqrt{k+n}+  \sqrt{k-n}\ ,\ \ \ \ L_2= \sqrt{k+n}-  \sqrt{k-n}\
$$

Two special cases are  $n=0$ 
and $n=k$. 

For $n=k$, one has $L_1=L_2$ and
 the classical configuration is a string whose projection in 
both the $Z_{1,2}$ planes
is  a circle. 
%This represents the configuration of maximal surface bounded by the string. 
This circular string rotates around its center like a wheel, so that the
classical distribution of string matter is stationary.

For $n=0$ the solution becomes a  straight rotating string.
Indeed, the resulting solution 
\beq
n=0:\ \ \ \ Z_1=  Z_2= 2\sqrt{k } e^{2i\tau } \ \cos(2\s)\ ,
\eeq
is classically equivalent to the string rotating in one-plane,
as is clear in the rotated frame
$Z_1'= 2\sqrt{2k } e^{2i\tau } \ \cos(2\s), \
Z_2'=0$ (cf. eq. (\ref{regg})).

However, at quantum level, the state 
\beq
|\Phi_{k,0}\rangle =  {1\over (k!)^2 } \ (b^\dagger )^{k}   (c^\dagger )^{k}
(\tilde b^\dagger )^{k}   (\tilde c^\dagger )^{k} |0\rangle \ ,
\label{zeze}
\eeq
is physically inequivalent from the state 
\beq
|\Phi_{k}^{J_{max}}\rangle  =  {1\over (2k)! } \ (b^{'\dagger} )^{2k}
(\tilde b^{'\dagger} )^{2k}|0\rangle \ ,
\label{unpla}
\eeq
representing a string rotating in a single plane, 
with $N_R=N_L=2k$ and maximal angular momentum, which we call $J_{max}$ state. 

Indeed, the state (\ref{unpla}) is an eigenstate of 
$J^2=\sum_{1\leq i<j\leq 9}(J_{ijR}+J_{ijL})^2$
with eigenvalue $J^2=16k^2+28k$ (maximal angular momentum 
for the given mass).
The states (\ref{ourstate}) however are not eigenstates of $J^2$. In fact
\beq
J^2  |\Phi _{k,n} \rangle = (12k^2-4n^2+28k)|\Phi _{k,n} \rangle 
+2(k-n)^2|\Phi_{k,n+1}\rangle +2(k+n)^2|\Phi_{k,n-1}\rangle
\eeq
%with \beq
%|other \rangle =\{2(k-n)^2(b^\dagger /\tilde b^\dagger )
%(\tilde c^\dagger /c^\dagger )+
%2(k+n)^2(\tilde b^\dagger /b^\dagger )(c^\dagger /\tilde c^\dagger
%)\}|\Phi _n \rangle \eeq
%$$\langle \Phi_n |other \rangle =0$$
%Moreover $\langle\Phi_n |J^2|\Phi _n \rangle =12k^2-4n^2+14k$ 
%which is less than the $J^2$-eigenvalue of the state 
%(\ref{unpla}) even for $n=0$.
In the large $k$ limit, the states $|\Phi _{k,n} \rangle , \ |\Phi_{k,n\pm
  1 }\rangle$ represent essentially the same semiclassical solution
(having the same values of $E$, $J_{12}$, $J_{34}$, and approximately
  the same values of $J_{12R}$, $J_{12L}$ and $J_{34R},\ J_{34L}$)
so effectively 
$J^2\cong (12k^2-4n^2+28k)+2(k-n)^2+2(k+n)^2=16k^2+28k\cong 16k^2$, 
which agrees with the value of $J^2$ of the
semiclassical solution.\footnote{
Note that all the states (\ref{ourstate}) and (\ref{unpla}) 
are eigenstates of $J_{R}^2$ and 
$J_{L}^2$ with the same eigenvalue $=4k^2+14k$; in the interactions, 
however, the conserved
operator is $J^2$.} 

\vskip0.5truecm

In the next section we will find   that the decay rates of the
state $|\Phi_{k,0}\rangle $ (corresponding to $n=0$) share
some qualitative features
with the decay rates for the state  $|\Phi_k^{J_{max}}\rangle $ (\ref{unpla})
computed in \cite{IR1},
though the quantitative  details of the quantum decay are different.

%%%%%%%%%%%%%%%%%%%%%%%%%%%%%%%%%%%%%%%%%%%%%%%%%%%%%%%%%%%
\subsection{Classically unbreakable closed strings}
%%%%%%%%%%%%%%%%%%%%%%%%%%%%%%%%%%%%%%%%%%%%%%%%%%%%%%%%%%%

An important feature of these solutions is that for $n\neq 0$ the
 closed string
cannot break classically in type II superstring theory. 
This splitting process can only happen if during the evolution
there is a time $\tau_0$ where  two points of the string get in contact, i.e. there are two
values $\s_1, \s_2$ such that $X^\mu (\s _1,\tau_0)=X^\mu
(\s_2,\tau_0)$.
For the straight  rotating (folded) closed string (\ref{regg}), the breaking can occur
at any time. In fact, it was found in \cite{IR1,IR2} 
that the string can decay
into two strings of masses $M_1,M_2$, where $M_2$ is a function of $M_1$. 
This relation between
$M_1,M_2$ which emerges in the quantum calculation
can be understood (and, in fact, accurately described) 
in terms of the semiclassical process of splitting.

When the classical breaking is not possible, one expects that the
decay channel into two large masses $M_1, M_2$ will be exponentially
suppressed (one can interpret this by saying that
breaking is possible only by tunnelling effect).
The reason is that for large masses $M,M_1,M_2$ each string in the
process should have a soliton semiclassical description, so that
the classical approximation is expected to apply, modulo terms which are
exponentially small in the masses. 
The original string can nevertheless
decay by emitting {\it light}  particles, 
for which  a semiclassical
soliton description does not apply.

This intuition will be confirmed by the quantum calculation in the
next section.
Moreover, in the cases we consider, we will find that only the rate of
{\it massless} particle  emission 
is not suppressed, with a rate that decreases as
the mass increases.
The conclusion would be that the lifetime of quantum states
associated with general classically unbreakable states should always be
very large, since only decay into massless particles is relevant.

\bigskip

It should be noted that there are numerous classical string solutions
where the classical breaking is not possible.
In particular, one can consider small perturbations of the
rotating ellipse solutions,
\beq
Z^1=Z^1_{\rm ellipse} +\epsilon (z_{1R}(\s^+)+z_{1L}(\s^-)) ,\ \ \ 
Z^2=Z^2_{\rm ellipse}+ \epsilon (z_{2R}(\s^+)+z_{2L}(\s^-)) ,
\label{fluctu}
\eeq
etc., where $Z^{1,2}_{\rm ellipse}$ is the solution 
given in (\ref{esr}) and $\epsilon\ll 1$. 
Starting with a given ``unbreakable'' classical
solution one can construct the corresponding quantum state.
{}For example, one may consider the solutions 
$$
Z_1= 
e^{2i\tau } \big( \sqrt{k+n}\  e^{-2i\s } + \sqrt{k-n'}\  e^{2i \s }
\big)\ ,
\ \ \ \ \ 
Z_2=e^{2i\tau } \big( \sqrt{k-n}\  e^{-2i\s } + \sqrt{k+n'}\  e^{2i \s
} \big)\ ,
$$
which cannot break,  indicating that the corresponding
quantum states with the same values of $J_{12R}, J_{12L}$
and $J_{34R}, J_{34L}$, 
\beq
|\Phi _{k,n,n'} \rangle ={\cal N}\ 
(b^\dagger )^{k+n}   
(c^\dagger )^{k-n} (\tilde b^\dagger )^{k-n'}   
(\tilde c^\dagger )^{k+n'} |0\rangle \ , 
\label{morestate}
\eeq 
$$
{\cal N}=\big[ (k+n)! (k-n)!(k+n')! (k-n')! \big]^{-1/2}
$$
should be long-lived.\footnote{
The classical solution is related  to (\ref{esr}) by an $O(4)$ rotation,  
but the quantum states $ |\Phi _{k,n,n'} \rangle$ and  
$ |\Phi _{k,n}\rangle$ are inequivalent.}

%%%%%%%%%%%%%%%%%%%%%%%%%%%%%%%%%%%%%%%%%%%%%%%%%%%%%%%%%%%%%
\section{Computation of ${\rm Im}(\Delta M^2)$ } 

%%%%%%%%%%%%%%%%%%%%%%%%%%%%%%%%%%%%%%%%%%%%%%%%%%%%%%%%%%%%%

We consider type II Superstring Theory in ten flat uncompactified dimensions.. 
The expression for the one-loop mass shift for the state 
$|\Phi_{k,n}\rangle $, 
with square mass $M^2_{k,n}=N=2k-1$ in units $\alpha{'}=4$, is 
(see Appendix C):
\be
&& \Delta M^2_{k,n}=c' \ g_s^2\ \int {d^2\tau\ov\tau_2^3}
\int d^2z \ e^{-4N{\pi y^2\ov\tau_2}}
\left| {\theta_1 (z|\tau )\ov\theta_1^{'}(0|\tau)}\right| ^{4N}
\big( {\pi\ov\tau_2} \big) ^{2N} \nonumber\\ 
&&\times \sum_{m_1,m_2} \big( {\pi\ov\tau_2} \big) ^{-m_1-m_2}
Q(k,n;m_1,m_2) \big( \p^2 \log(\theta_1(z|\tau ) \big)^{m_1} 
\big( \bar\p^2 
\log(\theta_1(z|\tau ) \big)^{m_2}\nonumber \\
\ee
where the numerical coefficient $Q(k,n;m_1,m_2)$ has been defined in
the 
Appendix C, and 
$c'$ is an overall constant normalization coefficient, which is independent of $k$ and $n$.
It can be seen that this expression is modular invariant.

In order to evaluate the imaginary part of $\Delta M^2_{k,n}$ we expand the 
holomorphic (and antiholomorphic) factors
in powers of $e^{i2\pi\tau}$ and $e^{i2\pi z}$ 
(respectively of $e^{-i2\pi\bar\tau}$ and $e^{-i2\pi\bar z}$).

It is convenient to define the coefficients $\gamma (m,p,q)$ by the following expansion:
\beq
(e^{i2\pi z})^N \left( 
{2\pi\theta_1 (z|\tau )\ov\theta_1^{'}(0|\tau)}\right) ^{2N}
\big( {1\ov 4\pi^2}\p^2 \log(\theta_1(z|\tau ) \big) ^{m_1}
= \sum_{p,q}\gamma (N,m_1;p,q)\ e^{i2p\pi\tau}e^{i2(q-p)\pi z}
\eeq
and similarly for the antiholomorphic coefficient 
$\gamma (N,m_2;\tilde p,\tilde q)$.

We have  to evaluate the imaginary part of the integral
\be
H &\equiv &\int {d^2\tau\ov\tau_2^{3+2N-m_1-m_2}}\ e^{i2(p-\tilde p)
\pi\tau_1 -2(p+\tilde p)\pi\tau_2}\nonumber\\
&\times & \int 
d^2z\ e^{-4N{\pi y^2\ov\tau_2}-2(q-p+\tilde q-\tilde p-2N)\pi y
+i2(q-p-\tilde q+\tilde p)\pi x}
\ee
The integrations over $\tau_1$ and over $x$ set $p=\tilde p$ and
$q=\tilde q$.
Then we use the formula
\be
&& {\rm Im} (H)={\rm Im} 
\int {d\tau_2\ov\tau_2^{3+2N-m_1-m_2}}
e^{-4p\pi\tau_2}
\int dy \ e^{-4N{\pi y^2\ov\tau_2}-4(q-p-N)\pi y}\nonumber\\
&&= c'' \ {(N\omega )^{2N-m_1-m_2+3/2}\ov \sqrt{N}\Gamma
  (2N-m_1-m_2+5/2)}\ ,
\label{imagy}
\ee
where $c''$ is a numerical constant independent of $N,p,q,m_1,m_2$ and 
$$
\omega =1-2({p\ov N}+{q\ov N})+({p\ov N}-{q\ov N})^2\ ,
$$ 
is the typical phase-space function. 
This can be seen by comparing the integrand in eq.(\ref{imagy}) with the
Schwinger parametrization of
a Feynman one-loop diagram for a field theory with vertex
$\sim \Phi\ \phi_1\phi_2$, 
with $\Phi$ representing the field of mass
$M$, and   $\phi_{1,2}$ the fields of masses $M_{1,2}$.
Then one sees that the integral leads to the one-loop correction $\Delta M^2$ 
due to the process $\Phi\to\phi_1 +\phi_2$ with 
$$
M_1^2=q \ \  and \ \ M_2^2=p \ .
$$
The number $2N-m_1-m_2-2$ is related to the orbital angular momentum carried
in the interaction (the $M_{1,2}$ particles having in general
intrinsic spins).

The function
$\omega =1-2({M_1^2\ov M^2}+{M_2^2\ov M^2})+({M_1^2\ov M^2}-{M_2^2\ov M^2})^2$
vanishes at the border of phase space $M_1+M_2=M$.

In conclusion, we get for the channel $q=M_1^2,p=M_2^2$
\be
{\rm Im} \Delta M^2_{k,n}(p,q) &=& c_0\  g_s^2\ \sum_{m_1,m_2}Q(k,n;m_1,m_2)
\gamma (m_1,p,q)\gamma (m_2,p,q)
\nonumber\\
& \times &
{N^{2N-m_1-m_2+1}\omega^{2N-m_1-m_2+3/2}\ov 4^{2N-m_1-m_2}\Gamma 
(2N-m_1-m_2+{5\over 2})}\ ,
\label{finra}
\ee
where $c_0$ is a numerical constant independent of $k,n,p,q$, which we
conventionally take as $c_0=32 (2\pi)^3$ . 
We also recall  $M^2=N=2k-1$ ($\alpha^{'}=4$ units). 

The decay rate for a given channel into particles of masses $M_1$
and $M_2$ is given by 
\beq
R_{k,n}(p,q)={ {\rm Im} \Delta M^2_{k,n}(p,q)\ov 2\sqrt{N}} \ .
\label{rata}
\eeq
This includes the contributions from all final states
with the same masses $M_1,M_2$.

The total decay rate is $R_{k,n}^{\rm total}=\sum_{p,q}R_{k,n}(p,q)$ and 
the lifetime of the Superstring state $|\Phi_{k,n} \rangle $ is 
\def\T{ {\cal T}}
\beq
\T_{k,n}= {1\over R_{k,n}^{\rm total}}\ .
\label{vida}
\eeq 

A final comment. Both $Q(k,n;m_1,m_2)$ and $\gamma (m,p,q)$ are integers 
which become very large in absolute value by increasing $k$. They have been 
dealt with by means of computer programs which manipulate integers without
approximations. In this sense we have performed an exact computation. 
The limitation comes from the machine size and we stopped our computation
at mass $M^2=129$ for the most stable state and $M^2=99$ for the others.
We will see that this appears to be enough for determining the asymptotic
(power-law) formula of the lifetime.

%%%%%%%%%%%%%%%%%%%%%%%%%%%%%%%%%%%%%%%%%%%%
\section{Decay rates and lifetimes}
%%%%%%%%%%%%%%%%%%%%%%%%%%%%%%%%%%%%%%%%%%%%
%\section{Numerical analysis.}
%%%%%%%%%%%%%%%%%%%%%%%%%%%%%%%%%%%%%%%%%%%%

We report here the main results for the decay $M\to M_1+M_2$.

(Conventions: $ c_0= 32 (2\pi)^3\ ,\  \alpha'=4$ ).

\begin{table}[h]\caption[Decay rates: Sum over massive channels 
$M_{1,2} \neq 0$ ]{Decay rates: Sum over massive channels 
$M_{1,2} \neq 0$ }
\label{bulkres}
%\begin{center}
\begin{tabular*}{13cm}{||@{\extracolsep{1.43cm}}c|@{\extracolsep{1.43cm}}c|
@{\extracolsep{1.43cm}}c|@{\extracolsep{1.43cm}}c|@{\extracolsep{1.43cm}}c||}
\hline
$M^2$  & $|\Phi^{J_{max}}\rangle$ & $|\Phi_{k, 0}\rangle$
    &$|\Phi_{k, \frac{k}{2}}\rangle$ & $|\Phi_{k, k}\rangle$  \\ \hline
\hline
$19$ & $3.450$ & $2.353$ & $0.373$ & $5*10^{-5}$ \\ \hline
$39$ & $4.713$ & $2.317$ & $0.097$ & $1.19*10^{-8}$ \\ \hline
$59$ & $5.321$ & $2.183$ & $0.033$ & $2.26*10^{-11}$ \\ \hline
$79$ & $5.718$ & $2.070$ & $0.014$ & $1.15*10^{-13}$ \\ \hline
$99$ & $6.013$ & $1.978$ & $0.007$ & $1.08*10^{-15}$ \\ \hline
$119$ & $~$     & $~$     & $~$     & $1.6*10^{-17}$\\ \hline
$129$ & $~$     & $~$     & $~$     & $2.2*10^{-18}$\\ \hline
\end{tabular*}%\end{center}
\end{table}

\begin{table}\caption[Decay rates: Sum over radiation channels $M_{1}$ or 
$M_2= 0$]{Decay rates: Sum over radiation channels $M_{1}$ or 
$M_2= 0$}
\label{una0res}
%\begin{center}
\begin{tabular*}{12cm}{||@{\extracolsep{1.42cm}}c|@{\extracolsep{1.42cm}}c|
@{\extracolsep{1.42cm}}c|@{\extracolsep{1.42cm}}c|@{\extracolsep{1.42cm}}c||}
\hline
$M^2$  & $|\Phi^{J_{max}}\rangle$ & $|\Phi_{k, 0}\rangle$
    &$|\Phi_{k, \frac{k}{2}}\rangle$ & $|\Phi_{k, k}\rangle$  \\ \hline
\hline
$19$ & $7.83$ & $7.02$ & $2.91$ & $0.249$ \\ \hline
$39$ & $6.19$ & $4.31$ & $1.27$ & $0.050$ \\ \hline
$59$ & $5.46$ & $3.32$ & $0.77$ & $0.019$ \\ \hline
$79$ & $5.02$ & $2.78$ & $0.53$ & $0.010$ \\ \hline
$99$ & $4.70$ & $2.44$ & $0.39$ & $0.0056$ \\ \hline
$119$ & $~$     & $~$     & $~$     & $0.0036$\\ \hline
$129$ & $~$     & $~$     & $~$     & $0.003$\\ \hline

\end{tabular*}%\end{center}
\end{table}

\begin{table}\caption{Lifetimes }
\label{lftmres}
%\begin{center}
\begin{tabular*}{12.3cm}{||@{\extracolsep{1.42cm}}c|@{\extracolsep{1.42cm}}c|
@{\extracolsep{1.42cm}}c|@{\extracolsep{1.42cm}}c|@{\extracolsep{1.42cm}}c||}
\hline
$M^2$  & $|\Phi^{J_{max}}\rangle$ & $|\Phi_{k, 0}\rangle$
    & $|\Phi_{k, \frac{k}{2}}\rangle$ & $|\Phi_{k, k}\rangle$  \\ \hline
\hline
$19$ & $0.0886$ & $0.107$ & $0.304$ & $4.010$ \\ \hline
$39$ & $0.0917$ & $0.151$ & $0.729$ & $19.830$ \\ \hline
$59$ & $0.0927$ & $0.181$ & $1.246$ & $51.912$ \\ \hline
$79$ & $0.0931$ & $0.206$ & $1.840$ & $103.711$ \\ \hline
$89$ &  $~$   &  $~$   & $~$    & $137.934$ \\ \hline
$99$ & $0.0933$ & $0.227$ & $2.510$ & $178.155$ \\ \hline
$109$ &  $~$   &  $~$    &   $~$     & $224.689$ \\ \hline
$119$ & $~$     & $~$     & $~$     & $277,834$\\ \hline
$129$ & $~$     & $~$     & $~$     & $337.877$\\ \hline

\end{tabular*}%\end{center}
\end{table}

\vskip0.5truecm

%%%%%%%%%%%%%%%%%%%%%%%%%%%%%%%%%%%%
\subsection{ Case $n\neq 0$ }
%%%%%%%%%%%%%%%%%%%%%%%%%%%%%%%%%%%%

 We have considered explicitly the maximal $n=k$ case and 
   the case $n=k/2$ (the latter is expected to illustrate 
   a generic case $0<n<k$,  with $n/k$   not $n/k\ll 1$).

In agreement with the fact that in this case the string does 
not break classically, 
the rate in the channels with both $M_1, M_2\neq0$ ($massive~ channels$) is much suppressed 
(for $n=k$, the ``maximally unbreakable'' string, it is completely negligible) 
compared to the channel in which $M_1$ or $M_2$ is equal to zero ($radiation~ channels$), 
see Tables \ref{bulkres} and \ref{una0res}.
{} In particular, from Table \ref{bulkres},  one observes that the sum of the 
decay rates for all the
channels with both $M_{1,2}\neq 0$ goes very rapidly
(exponentially) to zero.

Therefore the dominant decay mode is by emission of a massless particle (Table \ref{una0res}).

Figure 1 shows the decay rate  for $M_1=0$ as a 
function of $M_2^2$, in the case $n=k=40$.
We see that it is strongly peaked for $M^2_2=M^2-l$ with $l$ finite
and 
small with
respect to $M^2=2k-1$. 
Note that this figure gives the spectrum of the emitted massless
particle, since its energy is 
\beq
E=\frac{l}{2M}\ .
\eeq
 From this we see that the emitted massless particle is soft.

As for the other decay product, the massive one, we think that 
it is likely that 
it is classically similar to the decaying state, although 
quantum mechanically it can well be different (see also Appendix 
\ref{operatorialdecay}).

\vskip0.2truecm
\begin{figure}[h!]
\label{fig1} 
\vskip -0.5cm \hskip -1cm
\centerline{\epsfig{figure=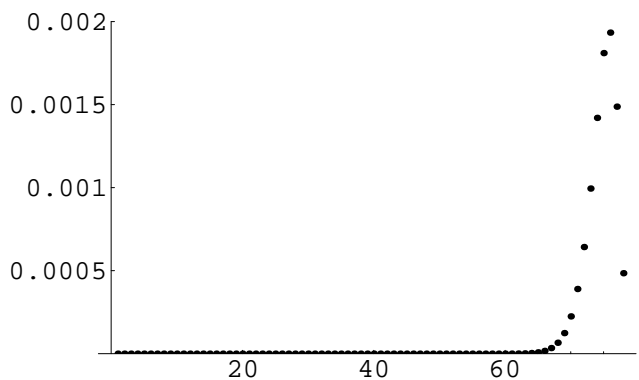,height=5.2truecm}}
\caption[Spectrum of massless particle emission for $n=k=40$.]{\footnotesize
Spectrum of massless particle emission for $n=k=40$,$M^2=79$. }
\end{figure}

%%%%%%%%%%%%%%%%%%%%%%%%%%%%%%%%%%
\subsection{ Case $n=0$}

In this case the rates for the channels $M_{1,2}\neq 0$ 
($massive~ channels$) and for the channels 
where $M_1$ or $M_2$ is zero ($radiation~ channels$) are comparable.

The massive channels $M_{1,2}\neq 0$ contain the case of classical breaking. 
In fact we have seen in Section 2 that $n=0$ is the limiting case where 
the elliptic classical configuration degenerates into a straight line:
 the classical closed string becomes a folded rotating string which can
break at any time.

In ref.~[12], by the description of the classical process of the
splitting 
into two closed
strings, we obtained the following masses of the decay products:

\be
M_1(a)&=& M\ \sqrt{ a^2+ {\sin^2(\pi a)\over \pi^2} }\ ,
\nonumber\\
M_2(a)&=& M\ \sqrt{ (1-a)^2+ {\sin^2(\pi a)\over \pi^2} }\ ,
\label{brek}
\ee
where $a \in [0,1]$ is a parameter specifying the splitting
point
$\s_0=a\pi $.
Eq.(\ref{brek}) defines parametrically a line $M_1=M_1(M_2)$.

We expect that the classical breaking configuration 
with masses (\ref{brek}) corresponds to the channel 
of maximum rate (for $M_{1,2}\neq 0$). Indeed the quantum results for
the decay rates are maximal on a curve $M_1=M_1(M_2)$ in the plane $M_1,M_2$. 
As an illustration we plot in Figure 2 the logarithm of the decay rate 
divided by $2M^2$
as a function of $M_1^2$ for fixed $M_2^2=1,5,10,20$ at $M^2=99$.

%\
%\vskip.2truecm
\begin{figure}[hbt]
\label{fig2} 
\vskip -0.5cm \hskip -1cm
\centerline{\epsfig{figure=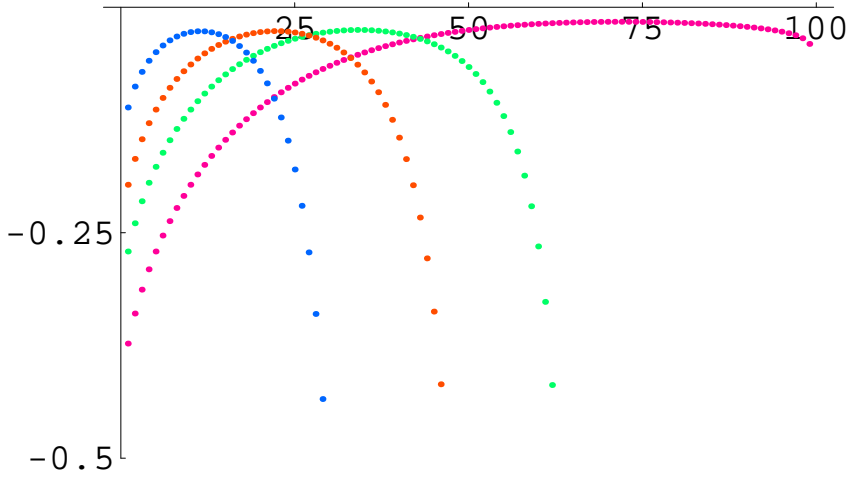,height=7truecm}}
\caption[Sections of the logarithm of the decay rate at fixed mass.]{\footnotesize  $M^2=99$,
logarithm of the decay rate divided by $2M^2$ as a function of $M_1^2$,
for  $M^2_2=$ \textcolor{magenta}{1} (flattest),\,\textcolor{green}{5},\,\textcolor{red}{10},\,\textcolor{blue}{20}
 (increasingly narrower). }
\end{figure}

Figure 3 displays the line of maximum value of the decay rate 
(in the plane $M_1/M,M_2/M$),
showing agreement with the classical curve (\ref{brek}).
This can be compared with the similar curve obtained in  \cite{IR1}
using saddle-point method, which applies at very large $M$ 
(see also \cite{IR2}).

%\footnote{
%In the comparison made in \cite{IR2} the classical curve accurately
%matches
%with the curve obtained from the one-loop quantum calculation. The
%reason of this accurate matching is that there the quantum curve
%was obtained by a saddle point evaluation of the quantum formula
%for ${\rm Im}\Delta M^2$, which amounts to taking the $M\to \infty $
%limit. 
%Whereas, here, we are exhibiting exact results for a given, 
%finite $M^2=2k-1$.}

\begin{figure}[hbt]
\label{fig3} 
\vskip -0.5cm \hskip -1cm
\centerline{\epsfig{figure=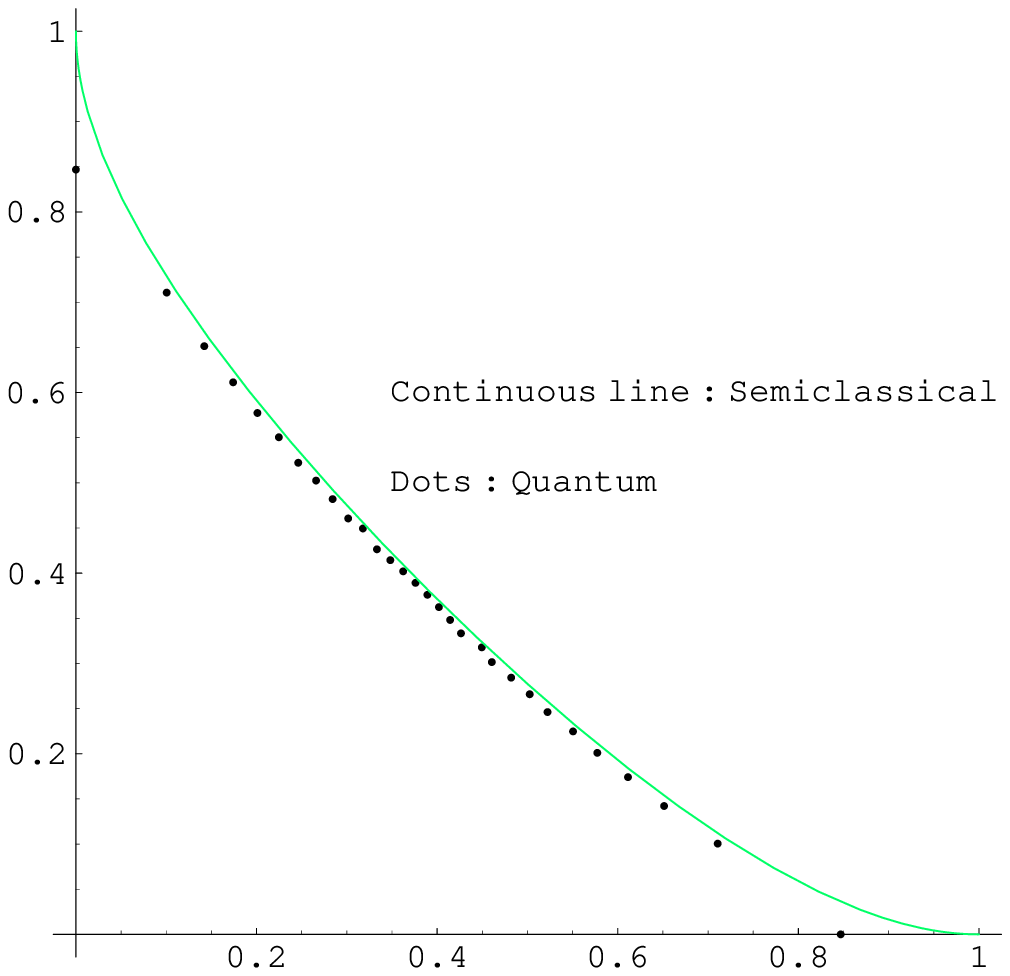,height=8truecm}}
\caption[Quantum $vs$ classical correlation of the masses of decay products.]{\footnotesize $M^2=99$.
 The continuum line corresponds to the classical 
$M_1=M_1(M_2)$ breaking curve (\ref{brek}). The dots are the values of
$M_1, M_2$ for which the rate of the quantum decay is maximum.\footnotemark}
\end{figure}

Near the end points of the classical curve ($M_1=0$ or $M_2=0$) the process 
of light or massless particles emission is not described by the classical string breaking.
Indeed we see in Figure 2 that the rates for $M_2^2=1$ are distributed in a rather flat way
and it is difficult to see a definite maximum.

The total rate for the sum of all the massive channels where both $M_{1,2}\neq 0$ is seen to
decrease with increasing $M$ (Table \ref{bulkres}, case $|\Phi_{k,0}\rangle$), but less fastly than 
the sum of the rates of the radiation channels with $M_1=0$ or $M_2=0$ 
(Table \ref{una0res}, case $|\Phi_{k,0}\rangle$).  
This is found quantitatively in Sect.4.3 by a fit of the data with a
power like formula.

Another interesting question is to investigate, among the different massive
channels with both $M_1$ and $M_2$ different from zero,
which ones gives the dominant contribution to the decay rate.
Figure 2 suggests that light-particle emission could be dominant 
over other massive
channels, and that the behavior  for the sum of all massive channels is
likely to result from a complex interplay of different individual behaviors,
the phase space features playing an important role.

\smallskip
For completeness we have reported in the Tables also the results for
the $J_{max}$ quantum state.

The three-dimensional plot of $S_0$ --~the logarithm of the decay rate 
divided by $2M^2$~-- as a
function of $M_1, M_2$ for the cases $n=0$ and $J_{max}$ looks 
qualitatively the same as the one for the  $J_{max}$ state reported in 
 Figure 4 of \cite{IR1}. 
One interesting question is if the maxima of 
the curve of Figure 3 approach $S_0=0$ at large $M$. This is the
expected result, since in the $n=0$ and $J_{max}$ cases 
the semiclassical breaking process should not be exponentially suppressed.
In a more detailed view we see that $S_0$ is rather 
smaller than zero for small $M$ and shows a number of structures
(see Appendix \ref{S0shape}): a peak
for $M_1=M_2$ and shallow maxima towards light values of $M_{1,2}$. 
By increasing $M$,
those structures tend to disappear. In the case $J_{max}$ the 
region of the maxima of $S_0$ 
get closer to zero, suggesting that $S_{0max} \rightarrow 0$ for  
$M\to\infty$.
In the case $n=0$, this is less evident.
%The three-dimensional plot of the logarithm of the decay rate as a
%function of $M_1, M_2$ for the case $n=0$ looks qualitatively the same
%as the one for the  $J_{max}$ state reported in \cite{IR1},
%so we do not reproduce it here.

In refs. \cite{IR1} and \cite{IR2}  the study of the $J_{max}$ state was
based on a saddle-point/WKB-type approximation. This gave an accurate 
determination of the exponential 
part of the formula for the decay rate, giving $S_{0max}=0$ 
in the $M\to\infty$ limit (precisely for the classical $M_1=M_1(M_2)$ 
curve (\ref{brek})~) 
but a less precise indication of the 
power behavior of the prefactor, 
and $\alpha =1$ was reported for the lifetime $\T\sim M^\alpha $. 
We will see in Section 4.3 that  the precise  power behavior of the lifetime 
as determined by the exact quantum calculation is different,
and that there is indeed a long-lived 
   rotating string state $|\Phi_{k, 0}\rangle$, although different at quantum level from
   the $J_{max}$ state.

%\footnote{
%It  should also be noted that  both   $|\Phi_{k,n=0} \rangle$ and
%$|\Phi^{J_{max}}\rangle$ states should describe the same rotating
%string soliton at
%large $k$, so it is conceivable that at very large $k$ they approach
%the same power behavior.}

%We see now that  a quantum version of the classical
%rotating string soliton is indeed long-lived although it is not precisely 
%the maximal angular momentum quantum eigenstate. 

\vskip0.5truecm

%The figures 1a and 1b show that the decay rate for given masses $M_1$,
%$M_2$  decreases as $n$ is increased,
%The exponential suppression of the decay rate
%into two large masses $M_1$, $M_2$ for the states with $n\neq 0$ is
%consistent with the interpretation in section 2 that the rotating ellipse
%cannot break classically. 

\vskip0.5truecm

%We have the lifetime (\ref{vida}) 
%for the cases $n=0, \ n=k/2, \ n=k$ (see fig. 4).
%In all cases we see that the lifetime increases with the mass.
%We find that for large masses it has a power-law dependence
%on the mass (in particular, this can be checked by  plotting 
%of  the logarithm of the
%lifetime in terms of the logarithm of the mass and checking that it
%is a straight line).

%The most stable state is the state with $n=k$.
%In this case, the lifetime rapidly increases with the mass and
%it is well fitted
%with the function
%$$
%\T _{k,n=k}={\rm const.}\ g_s^{-2}\  M^\alpha\ ,\ \ \ \ \alpha\cong 4.9
%$$

%\vskip0.5truecm

\footnotetext[4]{Since the maximum rate for $M_1=M_2$ occurs at $M_1^2=14$, we plot the points 
where $M_2$ gives the maximum for $0\leq M_1^2\leq 14$ and $viceversa$, for a total of $29$ points.}

%%%%%%%%%%%%%%%%%%%%%%%%%%%%%%%%%%%%%%%%%%%%%%%%%%%%%%%%%%
\subsection{Analysis of the results}\label{lftmbehav}
%%%%%%%%%%%%%%%%%%%%%%%%%%%%%%%%%%%%%%%%%%%%%%%%%%%%%%%%%%

To analize in a more detailed way the dominant contribution
to  the total decay rate, 
it is convenient to study the different channels separately.
The reason is that the decay rate in the channel in which a 
massless particle is emitted has a different
 behavior as compared to the channel in which two massive
particles are emitted. Considering the cases separately allows
for a more accurate convergence to the correct trend for $M\to \infty
$. [This separation is relevant only in the $n=0$
and $J_{max}$ cases, where the decay rate to two massive particles
is significant].

\smallskip

We fit the decay rates with a power-law behaviour, except for the massive channels 
in the cases $|\Phi_{k, k/2}\rangle$, $|\Phi_{k, k}\rangle$ which go very fast to zero. 

 To account for subleading corrections, we have fitted the decay rates 
assuming a dependence on the mass $M$ as follows:
\beq 
 f_r^{(1)}(M)=c\ M^{-\alpha}e^{\frac{\beta}{M}}\approx c M^{-\alpha}(1+\beta/M)
\eeq
 or 
\beq
f_r^{(2)}(M)=c\ M^{-\alpha}e^{\frac{\beta}{M^2}}\approx c M^{-\alpha}(1+\beta/M^2).
\eeq
We considered a favoured fit the one with minimum square deviation. 
The analysis has been carried through considering the logarithms of the data, 
and we take as the minimum square deviation $v^2\equiv \sum_i 
(\log[r(M_i)]-\log[f_r(M_i)])^2$, 
where the appropriate $r(M_i)$ can be read from Tables \ref{bulkres},
\ref{una0res}.

\smallskip

\begin{figure}[hbt]
\label{fig4} 
\vskip -0.5cm \hskip -1cm
\centerline{\epsfig{figure=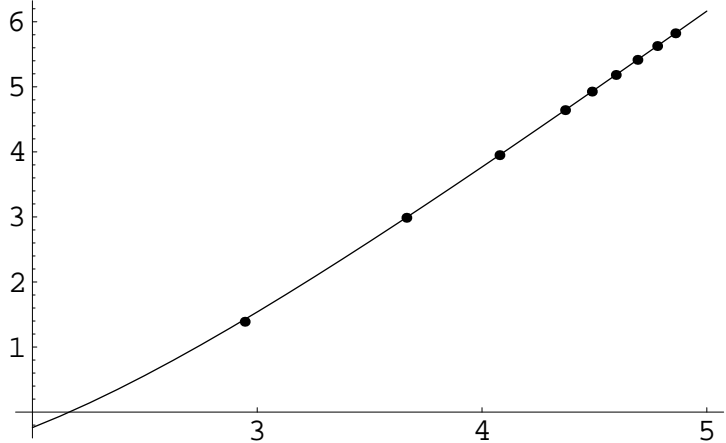,height=6truecm}}
\caption[Mass dependence of the logarithm of lifetime, case $n=k$.]{\footnotesize
The dots represent the logarithm  of the lifetime in the $n=k$ (``circular'')
case vs. the logarithm of $M^2$, for the different values of  Table 3.
The continuous line is based on the fit 
$\log{\mathcal{T}}=\frac{4.98}{2}\log{M^2}+\frac{8.36}{M^2}-\mbox{\scriptsize{$6.35$}}$
(with $v^2=\mbox{\scriptsize{$4\cdot 10^{-10}$}}$), see also (\ref{emedos}).}
\end{figure}

\smallskip

a) In the $J_{max}$ case (rotating string in one plane, maximal angular momentum) 
the decay is dominated by the sum of massive channels, 
containing the classical breaking. This is most favourably fitted by
\beq
\Gamma_{massive}=c \ g_s^2\ M^{0.25}e^{\frac{\beta}{M^2}}\ .
\eeq
with $v^2=2\cdot 10^{-6}$ (or $\Gamma_{massive}=c \ g_s^2\ M^{-0.14}e^{\frac{\beta}{M}}$, $v^2=2\cdot 10^{-5}$).

We get two comparable fits for the decay rate for the total of 
radiation channels, where at least one of the decay product is massless:
\be
&& \Gamma_{0}=c \ g_s^2\ M^{-0.54}e^{\frac{\beta}{M^2}} \ \ or \nonumber \\
&& \Gamma_{0}=c\ g_s^2\ M^{-0.46}e^{\frac{\beta}{M}}
\ee
with $v^2=8\cdot 10^{-8}$ in both cases.

\smallskip

b) In the case $n=0$
the sum of the massive channels
(which contains the classical breaking) dominates the decay, with a fit
\beq
\Gamma_{massive}=c \ g_s^2\ M^{-0.58}e^{\frac{\beta}{M^2}}\ ,
\eeq
with $v^2=2\cdot 10^{-6}$ (or $\Gamma_{massive}=c \ g_s^2\ M^{-0.90}e^{\frac{\beta}{M}}$, $v^2=2\cdot 10^{-5}$).

The decay rate for the total of radiation channels is fitted by:
\beq
\Gamma_{0}=c\ g_s^2\  M^{-0.96}e^{\frac{\beta}{M}} 
\eeq
with $v^2=7\cdot 10^{-8}$ (or $\Gamma_{0}=c \ g_s^2\ M^{-1.10}e^{\frac{\beta}{M^2}}$, $v^2=9\cdot 10^{-7}$).

\smallskip

c) For the case $n=k/2$, as already mentioned,  
the sum of the radiation channels, in which $M_1$ or $M_2$ is equal to zero, 
is the by far dominant contribution,
with a fit 
\beq
\Gamma_0=c \ g_s^2\ M^{-3.02}e^{\frac{\beta}{M}}\ ,
\eeq
with $v^2=4\cdot 10^{-5}$ 
($\Gamma_0=g_s^2\cdot c\cdot M^{-2.74}e^{\frac{\beta}{M^2}}$ 
gives $v^2=7\cdot 10^{-5}$).

\smallskip

d) For the case $n=k$ the sum over the radiation channels utterly predominates 
(the massive channels contribution is absolutely negligible).
In this case we have more data for the lifetime given in Table 3.
Since we are interested in the asymptotic behavior, 
for the fit we
only use the largest values of $M^2=79,89,99,109,119,129$.
We obtain
\beq
\Gamma_0=c\ g_s^2\  M^{-4.98}e^{\frac{\beta}{M^2}}
\label{emedos}
\eeq
with $v^2=6\cdot 10^{-10}$ ($\Gamma_0=c\ g_s^2\ M^{-5.1}
e^{\frac{\beta}{M}}$ gives $v^2=8\cdot 10^{-10}$).

\smallskip

 Finally, from the above analysis, we compute the lifetime behavior 
with the mass $M$, 
that is the one given by the predominant decay channels only, by:
\beq
\T =1/\Gamma(dominant)\ .
\eeq
 This gives the lifetime behavior for our states in the limit $M\to \infty$. 
Notice that in the cases $J_{max}$ and $n=0$ a naive fit to the total 
lifetime given in Table \ref{lftmres}
would not follow this result. In fact, the trend of the massive decay rate,
dominating over the rest, is disguised 
by the fact that the sum of the radiation channels is of comparable magnitude. 

 In conclusion our results indicate, in the large $M$ limit, for the 
various cases, a power-like behaviour for the lifetime as:
\beq
\ \T=\frac{c}{g_s^2}\ M^{\alpha}\ .
\eeq
 In Table \ref{lftmpower} we summarize the values found for $\alpha$ in the various cases.
The uncertainty on $\alpha$ can be estimated by comparing different 
fits, as it has been
reported above. 
 For the most interesting ``circular'' case $n=k$, 
the value of $\alpha=4.98$ has a very small uncertainty, 
as it can also be seen from Figure 4. 
Also  the value of $\alpha=3.0 $ for the intermediate case $n=k/2$ has
a small uncertainty, although in this case the accuracy in the
determination of $\alpha $ is lower than in the $n=k$ case,
because the numerical results for $n=k/2$ are so far available only up to
$M^2=99$.
For the breakable cases, $n=0$ and $J_{max}$, the values of $\alpha $
given in the Table 4.3 have a larger uncertainty,
which is more difficult to estimate,
because, as we said, there are competing decay channels of the same 
magnitude but different behavior 
(and, moreover, neither in those cases we 
could  obtain numerical results beyond $M^2=99$).

\begin{table}[htbp]\label{lftmpower}
\centering
\begin{tabular}{||l|l||}
Cases & $\alpha $    \\
 \hline
$J_{max}$ & -0.25    \\
$n=0$ & 0.58     \\
$n=k/2$ & 3.02    \\
$n=k$ & 4.98   \\ 

\end{tabular}
\parbox{5in}{\caption[Asymptotic lifetimes for the different cases.]{
Lifetimes for the different cases: $\T \sim M^{\alpha}$. See the discussion in the text for the uncertainty on $\alpha$.
}} \end{table}

\smallskip

In Appendix E we report a computation with the operator formalism 
giving the behavior with $M$ of the decay rate for the channel: 
\beq
state\{ n=k\}\to graviton ~+~ state\{ n=k-l \}
\label{chh}
\eeq
We obtain a lifetime for this mode  $\T\cong c(l)\ M^{5}$
for any $l$ small with respect to $k$ (that is to $M^2$).
This is very near the result $\alpha =4.98$, and suggests
that the channels $(M_1^2=0) ~+~(M_2^2=M^2-l)$ dominate 
the decay (see Fig. 1). 
Note that, a priori, the power
law behavior of the decay rate for this particular channel (\ref{chh})
needs not be
the same as the formula for the full decay rate,
since  there are many other states with $M_2^2=M^2-l$. 
%different from the ones selected in Appendix E.

%%%%%%%%%%%%%%%%%%%%%%%%%%%%%%%%%%%%%%%
\section{Discussion}
%%%%%%%%%%%%%%%%%%%%%%%%%%%%%%%%%%%%%%%%%

To summarize, the exact evaluation of the one-loop contribution 
to ${\rm Im}(\Delta M^2)$
shows 
that
there are massive states in type II superstring theory which
are almost stable, with a lifetime growing with the mass 
as fast as $\T = {\rm const.}\ g_s^{-2} \ M^5$.

We have seen that these states can decay only by masless emission
(other channels being exponentially suppressed), in agreement
with the interpretation  that they cannot break classically.
We repeat that these are superstring type II states. Indeed they do not exist  in type I superstring 
theory, due to the fact that the type I superstring is unoriented.

A natural question is how the decay rates computed here are 
corrected by higher loops.
One particular sector of higher loop corrections are
gravitational effects,
which  are expected to be of order $g_s^2M $.
%, and hence
%they should become important for sufficiently large masses.
Therefore they should not affect the formula for the 
lifetime $\T = {\rm const.}\ g_s^{-2} \ M^5$
as long as $M\ll 1/g_s^{2}$ (in units of $\alpha' $).
This indicates that 
the lifetime can be as large as $\T = {\rm const.}\ g_s^{-12}$ \ !

{}For states with masses larger than $1/g_s^{2}$, gravitational
effects can no longer be ignored. 
However, note that none of these states can 
become black holes at larger masses.
The reason is that the spatial extension of these states
grows linearly with the mass, i.e. $L\sim M/\a' $ (see sect. 2), 
whereas
in ten dimensions the Schwarzschild radius grows with the 
mass as $R_{\rm sch}\sim M^{1/7}$,
i.e. for large masses it is always the case that the size 
$L$ is much larger than $R_{\rm sch}$.
As $M$ is increased, the gravitational field near a segment
of the string becomes strong, and the question is what happens
to the string configuration at larger couplings.
In particular, the most stable state of the 
family, the $n=k$ state representing a rotating circular string,
seems a very robust string configuration and it is 
plausible that it may survive in
the strong coupling regime. 
%When $g^2M $ becomes large,
%it is conceivable that a horizon could form surrounding the string.
%In fact, 
This state (or a mixture of states associated with
small fluctuations (\ref{fluctu}) near this state)
is a natural candidate
for becoming a black ring \cite{ER} in the strong coupling regime.
The fact that this rotating $n=k$ string solution exists in $D=5$ but
not
in $D=4$ is consistent with the fact that there is no black ring in
four dimensions.
It would be interesting to study the correspondence principle
\cite{HP} for this case.

\medskip

A question of interest is how many states --among the exponential
number of states existing at each mass level-- are long-lived.
This is obviously a difficult question to answer in general, 
without an explicit computation case by case.
Nevertheless, one can try to estimate the number of classically unbreakable
states, since for such states the only relevant decay channel is by 
emission of light particles.  
Like in the examples discussed in this paper, 
they are expected to have a longer lifetime.

It should be noted that the higher is the dimension, the less is 
probable that two points
of the string get in contact (e.g. viewing the average string state 
as a random walk process).
This suggests that in ten dimensions, a type II superstring state has little chance to
break.  Therefore, for most states the decay would dominantly occur
   through energy leakage by radiation of massless (or light) modes 
(rather than decay into two very massive particles, which requires breaking 
of the string).

We have found that in the classically breakable $n=0$  case  the
lifetime grows as $\T\sim g_s^{-2}\ M^{0.58}$, and that this is determined
by the asymptotically dominant channel of string breaking.
 This would suggest
that a  {\it generic} closed string state should also become more stable 
for larger masses.
%have a longer lifetime than the $n=0$ case. 
Indeed, in the $n=0$ case, the closed string is folded
with all points in contact at all times, so it can break at any time.
Instead, a generic closed string solution 
can only break at a discrete set of specific times
and at the specific place  where two points touch. For this reason,  
the decay rate into two massive particles
of a generic state should be lower than in the $n=0$ case.
Of course, it is possible that for some states the 
decay channel by
massless particle emission is more relevant,
%(since many higher frequency oscillator modes  excited), 
with the effect of 
reducing the lifetime as compared to the $n=0$ case.

\smallskip

It seems unlikely 
that the family of states (\ref{ourstate}) we have considered here constitutes the only states
that survive the large mass limit (in particular, we expect that the states of the larger family (\ref{morestate})
are also long-lived). 
Long-lived states are very visible among the states produced 
in a high energy collision, and 
it is plausible  that there is a vast sector 
of long-lived states in type II superstring theory
which could be produced, i.e. a vast sector of states
whose lifetime increases with the mass. 
This is an exciting prospect which deserves further investigation.

%%%%%%%%%%%%%%%%%%%%%%%%%%%%%%%%%%%%%%%%%%%%%%%%%
\section*{Acknowledgments}
%%%%%%%%%%%%%%%%%%%%%%%%%%%%%%%%%%%%%%%%%%%%%%%%%%

We acknowledge partial support by
the European Community's Human potential
Programme under the contract HPRN-CT-2000-00131.
The work of J.R. is
supported in part also by MCYT FPA,
2001-3598 and CIRIT GC 2001SGR-00065.

\bigskip 

%%%%%%%%%%%%%%%%%%%%%%%%%%%%%%%%%%%%%%%%%%
\appendix
\section{Quantum state construction}
%%%%%%%%%%%%%%%%%%%%%%%%%%%%%%%%%%%%%%%%%%

As specified in section 2 we are interested in 
one particular family of superstring states. 
In eq. (\ref{bosonsol}) we introduced only the bosonic part 
of our states. Here we construct the full
 Superstring state.

Define $Z^1$ and $Z^2$ as in section 2, with the expansions
(\ref{zetu}), (\ref{zetdu}).
The bosonic contribution to the 
 Superstring state is as in eq. (\ref{ourstate}), which
we repeat for convenience:
\beq
|\Phi_{k,n}\rangle=a_n\ (b^\dagger)^{k+n}(c^\dagger)^{k-n}(\td b^\dagger)^{k-n}(\td c^\dagger)^{k+n}|0\rangle \label{bosonstat}
\eeq
where $a_n$ is a normalization factor:
\beq
a_n=\frac{1}{(k+n)!(k-n)!}
\eeq

%We recall section \ref{solitonstwopl} for justifying our 
%identification of these family of states to a family of string solitons.

%Note:
%il momento angolare si calcola facilmente a partire da 
%J^{\mu\nu}=T\int^\2pi_0 d\s 
%(X^\mu\frac{dX^\nu}{d\tau}-X^\nu\frac{dX^\mu}{d\tau})
%l'energia si computa a partire da
%P^0=T\int_0^2\pi  \partial_\tau X^0 per la stringa classica, dove X^0=2\sqrt{k}\tau, per soddisfare i constraints
% ovvero dal solito
%p_0^2=\Sum p_i^2+N+a
%per lo stato quantistico.

Now we proceed to the construction of  the superstate. 
We look for a state with the same values of energy and  angular momentum
components. 
%in order to respect the map string states/solitons we have introduced
%before. 
The  superstate satisfying these requirements
is given by:
\beq
|\Phi_{k,n} \rangle = \mathcal{N}a_n\ |\phi_n^R  \rangle  |\phi_n^L  \rangle
\label{complete}
\eeq
with
\beq
|\phi^R_n\rangle=
%\mathcal{N}a_n
\left(
(k+n)(\psi^{\rm z_1}_{-\half})^\dag c_{1}^\dag+(k-n)(\psi^{\rm z_2}_{-\half})^\dag
b_{1}^\dag
\right)
(b_{1}^\dag)^{k+n-1}(c_{1}^\dag)^{k-n-1}|0, p\rangle \ ,
\label{estado}
\eeq
where we have written  only the right-moving part 
(the left-moving is obtained by substituting all the 
oscillators with tilded operators and changing $n\rightarrow -n$). 
It is obtained from (\ref{bosonstat}) by applying the mode
$$
G_{1/2}=\sum_{n \epsilon \mathbb{Z}} \alpha^\mu_{n}\ \psi_{\mu\ 
  {1\over 2}-n} 
$$ 
of the superconformal current (the analog must be done 
in the left-moving sector). Here we have used real coordinate modes 
in order to simplify the  notation.

The state (\ref{estado}) satisfies all the physical constraints, 
namely (we write  only those of the right-moving sector)
\be
 & G_r| \Phi _{k,n} \rangle & =0\ , \ \ \ \ \ r\geq 1/2 \ ,\nonumber\\
 & L_n| \Phi _{k,n} \rangle & =0 \ , \ \ \ \ \ n\geq 1 \ ,\nonumber\\
 & (L_0-\half)| \Phi _{k,n} \rangle & =0\ ,
\ee
where $L_n, G_r$ are the super-Virasoro algebra 
generators and we are in the Neveu-Schwarz sector 
(the zero-point energy is equal to $\half$).

The normalization constant in front of the state can be easily
computed by requiring $\langle \Phi_{k,n} |\Phi_{k,n} \rangle =1$, giving:
\beq
\mathcal{N} = {\frac{1}{2k}}\ ,
\label{norm}
\eeq
i.e.  an additional factor ${1\ov 2k}$ with respect to the bosonic
state
(\ref{bosonstat}).

Finally, one can check that the  mass-squared operator $M^2$ acting on
this  state gives
\beq
\alpha ' M^2=4 (2k-1)\ .
\eeq

%%%%%%%%%%%%%%%%%%%%%%%%%%%%%%%%%%%%%%%%%%%%
\section{Vertex operators}
%%%%%%%%%%%%%%%%%%%%%%%%%%%%%%%%%%%%%%%%%%%%

Here we construct the vertex operators 
corresponding to the state we have introduced in the preceding
section. 
From now on we will consider explicitly only the right-moving
(holomorphic) part of
the 
complete vertex. The construction of the left-moving (anti-holomorphic) part 
is similar.

The vertex operator for a superstring state can be
constructed in different pictures. 
Following the definitions of \cite{polstr}, 
we shall use the ``integrated'' form of the vertex operator.

Given an operator  $W$  of conformal dimension $h_W=\half$, 
a vertex operator $V$  of conformal dimension
$h=1$ can be obtained by writing
\beq
V=G_{-\half}\cdot W \ ,
\eeq
where ``$A\cdot B$ '' indicates the usual commutator (or anticommutator)
of the operators $A, B$. \footnote{An equivalent way
 to construct a vertex operator is to define
$V=G_r\cdot W$ by requiring $V$ to be independent of $r$. We have
also pursued this way of computing $V$ finding that indeed it leads to the
same final expression.}

As operator $W$ we take
\def\z{{\rm z} }
\beq
W_{k,n}= \big( (k+n)\psi^{\z_1} \partial Z_2+(k-n)\psi^{\z_2} \partial
Z_1 \big) \partial 
{Z_1}^{k+n-1}\partial {Z_2}^{k-n-1}\, ({\rm rightmovers})e^{ip\cdot X},
\eeq
which has conformal dimension $h_W=\half$ provided $p^2=2k-1$.

Acting on this operator with the superconformal charge  produces a 
certain number of terms. 
We are interested only in the terms which will give a 
non-vanishing contribution to the one-loop amplitude.
As shown in appendix \ref{spinstruct}, 
only a few  terms with  a specific combination of fermionic operators 
give a non-zero contribution in the two-point function
on the torus.

Considering therefore only these terms, our final vertex operator is 
$$
V_{k,n}=\it{Norm} V^L V^R e^{ip\cdot X}\ ,
$$
with
\begin{eqnarray}
  V^R &=&(\psi^{\z_1}\partial \psi^{\z_2}+\psi^{\z_2}\partial \psi^{\z_1})
(k^2-n^2)(\partial Z_1)^{k+n-1}(\partial Z_2)^{k-n-1}  \\ \nonumber
 & +&(k+n)(k+n-1)\psi^{\z_1}
\partial\psi^{\z_1}(\partial Z_1)^{k+n-2}
(\partial Z_2)^{k-n} \\ \nonumber
 & +&(k-n)(k-n-1)\psi^{\z_2}\partial \psi^{\z_2}(\partial Z_1)^{k+n}
(\partial Z_2)^{k-n-2}
\ .
\end{eqnarray}
The {\it Left} part of the vertex $V^L_{k,n}$ is the same with $\p\rightarrow\bar\p$ and 
$\psi\rightarrow\tilde\psi$. 

We recall that the normalization constant is:
\beq
{\it Norm}=\frac{1}{2k(k+n)!(k-n)!}\left(\sqrt{\frac{2}{\alpha'}}\right)^{4k-4}
\eeq
In what follows  we choose units where $\alpha'=4$.

%%%%%%%%%%%%%%%%%%%%%%%%%%%%%%%%%%%%%%%%%%%%%%%%%%%%%%%%%
\section{Computation of the amplitude}\label{appampl}
%%%%%%%%%%%%%%%%%%%%%%%%%%%%%%%%%%%%%%%%%%%%%%%%%%%%%%%%%

The correlator $\langle \bar V_{k,n}V_{k,n}\rangle $ on the torus 
is (after summing over spin structures, 
see Appendix \ref{spinstruct}):
\beq
\langle \bar V_{k,n}(z)V_{k,n}(0)\rangle = 
{1\ov (2k(k+n)!(k-n)!)^2}
\langle  \mathbf{O}_{k,n}^R(z)\mathbf{O}_{k,n}^L(z) \rangle
\langle e^{-ipX(z)}e^{ipX(0)} \rangle 
\label{unve}
\eeq
where
\be
 \mathbf{O}_{k,n}^{R,L}(z) &=&
 4(k^2-n^2)^2 O^{R,L}_\alpha (z,0)+
(k+n)^2(k+n-1)^2 O^{R,L}_\beta (z,0)
\nonumber\\
&+&
(k-n)^2(k-n-1)^2 O^{R,L}_\gamma (z,0)\ ,
\ee
and
$$
 O^R_\alpha (z,0)=
(\p\bar Z_1)^{k+n-1}(z)(\p\bar Z_2)^{k-n-1}(z)(\p Z_1)^{k+n-1}(0)
(\p Z_2)^{k-n-1}(0) \ ,
$$
$$
 O^R_\beta (z,0)=
(\p\bar Z_1)^{k+n-2}(z)(\p\bar Z_2)^{k-n}(z)(\p Z_1)^{k+n-2}(0)
(\p Z_2)^{k-n}(0) \ ,
$$
$$
 O^R_\gamma (z,0)=
(\p\bar Z_1)^{k+n}(z)(\p\bar Z_2)^{k-n-2}(z)(\p Z_1)^{k+n}(0)
(\p Z_2)^{k-n-2}(0) \ ,
$$
and the  ``Left'' Operators $O^L$ are the same 
as the corresponding $O^R$ with $\p\rightarrow\bar\p$.  

We see that we have $9$ possible contractions $\langle O^RO^L \rangle $.
The basic correlators are 
\be
\langle
\p\bar Z_i(z)\p Z_j(0)\rangle
=-\delta_{ij}\ 2\left( 
\p^2  \log \theta_1(z|\tau )+{\pi\ov\tau_2}\right) \ ,\nonumber\\
\langle \bar\p\bar Z_i(z)\bar\p Z_j(0)\rangle =
-\delta_{ij}\ 2\left( \bar\p^2 \log \bar\theta_1(z|\tau
)+{\pi\ov\tau_2}\right)\ ,
\nonumber\\
\langle \bar\p\bar Z_i(z)\p Z_j(0)\rangle =
\langle \p\bar Z_i(z)\bar\p Z_j(0) \rangle
=\delta_{ij}2\ {\pi\ov\tau_2}\ ,  
\nonumber\\
\langle 
e^{-ipX(z)}e^{ipX(0)} \rangle 
=e^{-4(2k-1){\pi y^2\ov\tau_2}}\ 
\left| {\theta_1 (z|\tau )\ov\theta_1^{'} (0|\tau )}\right| ^{4(2k-1)}
\ee
(here $y={\rm Im} (z)$).
The generic term $\langle O^RO^L \rangle $ gives 
(with the appropriate $m,n,\bar m,\bar n$):
$$
\langle \ (\p\bar Z_i)^{m}(z)
(\p Z_i)^{n}(0)(\bar\p\bar Z_i)^{\bar{m}}(z)(\bar\p
Z_i)^{\bar{n}}(0)\ \rangle 
$$
\be
&=& \sum_j {m!n!\bar m!\bar n!\ov j!\bar j!(m-j)!(\bar m-\bar j)!}
\langle \p\bar Z_i(z)Z_i(0)\rangle ^j
\langle \bar\p\bar Z_i(z)\bar Z_i(0)\rangle ^{\bar j}\nonumber\\
&\times & \langle \p\bar Z_i(z)\bar\p Z_j(0)\rangle ^{m-j}
\langle \bar\p\bar Z_i(z)\p Z_j(0)\rangle ^{\bar m-\bar j}
\ee
Note that in every term  the relations 
$m-j=\bar n-\bar j, ~~\bar m-\bar j=n-j$ hold, 
and thus $\bar j=j-(m-\bar n)=j-(n-\bar m)$.
We finally obtain:
\be
\langle 
\bar V_{k,n}(z)V_{k,n}(0) \rangle  
&&= 
{(k+n)!^2(k-n)!^2\ov 4k^2}
\left| {\theta_1 (z|\tau )\ov\theta_1^{'} 
(0|\tau )}\right| ^{4(2k-1)} e^{-4(2k-1)\pi y^2\ov\tau_2}\nonumber\\
&&\times \sum_{j,l}  c(j,l;k,n)
\big( {\pi\ov\tau_2} \big) ^{2(2k-1-j-l)-2}
\left| \p^2 \log\theta_1(z|\tau )+{\tau_2\ov\pi}\right| ^{2(j+l)}\nonumber
\ee
The sum runs over $2n-2\leq j\leq k+n,0\leq l\leq k-n$, 
\beq
c(j,l;k,n)={\sum_{i=1}^9 S_i(j,l;k,n)\ov j!l!(k+n-j)!^2(k-n-l)!^2(j-2n+2)!(l+2n+2)!}
\eeq
and 
$$
\sum_{i=1}^3 S_i(j,l;k,n)=(k+n-j)^2(k-n-l)^2A(j,l;k,n)\ ,
$$
$$
\sum_{i=4}^6 S_i(j,l;k,n)=(k+n-j)^2(k+n-j-1)^2B(j,l;k,n)\ ,
$$
$$
\sum_{i=7}^9 S_i(j,l;k,n)=\sum_{i=4}^6 S_i(l,j;k,-n)\ ,
$$
with
\be
A(j,l;k,n) &=& 4(j-2n+2)(l+2n+2)(l+2n+1)(l+2n)
\nonumber\\
&+& 4(j-2n+2)(j-2n+1)(j-2n)(l+2n+2)
\nonumber\\
&+& 16(j-2n+2)(j-2n+1)(l+2n+2)(l+2n+1)\ ,
\ee
\be
B(j,l;k,n) &=& 4(j-2n+2)(l+2n+2)(l+2n+1)(l+2n)\nonumber\\
&+&(l+2n+2)(l+2n+1)(l+2n)(l+2n-1)\nonumber\\
&+& (j-2n+2)(j-2n+1)(l+2n+2)(l+2n+1)\ .
\ee

Note that for the particular case $n=k$ the above expressions
simplify, 
since the only nonvanishing
contribution occurs for $j=2k-2,l=0$ and $c(2k-2,0;k,k)=1/(2k-2)!^2$.

Expanding  the holomorphic and antiholomorphic binomial factors 
we obtain
\be
\langle 
\bar V_{k,n}(z)V_{k,n}(0) \rangle 
=e^{-4(2k-1){\pi y^2\ov\tau_2}} 
\left| 
{\theta_1 (z|\tau )\ov\theta_1^{'}(0|\tau)}
\right|^{4(2k-1)}
({\pi\ov\tau_2})^{2(2k-1)-2}\nonumber\\
\sum_{m_1,m_2}({\pi\ov\tau_2})^{-m_1-m_2}
Q(k,n;m_1,m_2)(\p^2 \log(\theta_1(z|\tau ))^{m_1}(\bar\p^2 
\log(\theta_1(z|\tau ))^{m_2}
\ee
where
\beq
Q(k,n;m_1,m_2)={(k+n)!^2(k-n)!^2\ov 4k^2}
\sum_{j,l} { c(j,l;k,n) (j+l)!^2\ov m_1!m_2!(j+l-m_1)!(j+l-m_2)!}
\eeq
{}In the particular case $n=k$ we get 
\beq
Q(k,k;m_1,m_2)={(2k-1)!^2\ov m_1!m_2!(2k-2-m_1)!(2k-2-m_2)!}\ .
\eeq
The computation for this $n=k$ case can be done 
directly from the vertex
\beq
V_{k,k}=\frac{1}{2^{2k-2}}
{(2k-1)^2\ov (2k-1)!^2} \psi _{\z_1} \p \psi _{\z_1} 
(\p Z_1)^{2k-2}
\tilde\psi_ {\z_2} \bar\p \tilde\psi _{\z_2}  
(\bar\p Z_2)^{2k-2} e^{ipX} \ ,
\eeq
noticing that in this case 
\beq
\langle \bar V_{k,k}(z)V_{k,k}(0)\rangle  
=(2k-1)^2 \langle  e^{-ipX(z)}e^{ipX(0)} \rangle
\left| \p^2 \log \theta_1(z|\tau )+{\pi\ov\tau_2} \right| ^{2(2k-2)} \ .
\eeq

\vskip0.5truecm

Another state of interest is the state of maximal angular 
momentum with $M^2=2k-1$ which rotates in one plane only. 
In this case the superstate is (cf. eq. (\ref{unpla}), for the bosonic version)
\beq
|\Phi_{k}^{Jmax} \rangle ={1\ov (2k-1)!}
{\psi _1}^{\dagger}(b^{\dagger})^{2k-1}
{\tilde\psi _1}^{\dagger}(\tilde b^{\dagger})^{2k-1}|0\rangle \ .
\eeq 

The corresponding vertex is given by 
\beq
V_k^{Jmax}=\frac{1}{2^{2k-2}}{(2k-1)^2\ov (2k-1)!}{\psi _{\z_1} }\p{\psi _{\z_1}} (\p Z_1)^{2k-2}
{\tilde\psi _{\z_1}}\bar\p{\tilde\psi _{\z_1}} (\bar\p Z_1)^{2k-2} e^{ipX}\ ,
\eeq
and
\be
&&\langle \bar V_{k}^{Jmax}(z)V_{k}^{Jmax}(0)\rangle =
\langle e^{-ipX(z}e^{ipX(0)} \rangle \nonumber\\
&\times & \sum_j {(2k-1)!^2\ov j!^2(2k-2-j)!^2}
\left| \p^2 \log \theta_1(z|\tau )+{\pi\ov\tau_2}\right| ^{2j}
\big( {\pi\ov\tau_2} \big) ^{2(2k-1)-2j-2}\ .
\ee
One obtains
\beq
Q^{Jmax}(k,m_1,m_2)=
{(2k-1)!^2(2(2k-1)-m_1-m_2)!\ov m_1!m_2!(2k-2-m_1)!^2(2k-2-m_2)!^2}\ .
\eeq

\medskip

{}In all cases $\Delta M^2$ is finally obtained by the integral:
\beq
\Delta M^2=\int {d^2\tau\ov\tau_2^5}\int d^2z 
\ \langle \bar V(z)V(0) \rangle 
\eeq
by inserting the appropriate vertex.

We will use convenient formulas for the expression involved in the computation 
of the amplitude (\cite{Okada,polstr})
\beq
  2\pi i \left(e^{2\pi iz}\right)^\half 
\frac{\theta_1(z, \tau)}{\theta^{'}_1(0, \tau)}= 
\eeq
 $$
 =- \left(\prod_{n=1}^\infty [1-(x_1 x_2)^n]^3\right)^{-1}
    \left( \sum_{n=1}^{\infty}(-1)^{n-1}x_1^{n(n-1)/2}x_2^{(n-1)(n-2)/2}
(x_2^{2n-1}-1)\right) 
$$

\vspace{0.5truecm}

\beq
 \frac{1}{4\pi^2}\partial^2_z(\log \theta_1(z, \tau))\, =\, 
 \sum_{n= 1}^\infty nx_2^n + \sum_{n, l= 1}^\infty
(x_1 x_2)^{nl}(x_2^n+x_2^{-n})
%\nonumber
\eeq
with $x_1=e^{2\pi i(\tau-z)}$, $x_2=e^{2\pi iz}$.

%%%%%%%%%%%%%%%%%%%%%%%%%%%%%%%%%%%%%%%%%%%%%%%%%%%%%%%%%%%%
\section{Sum over spin structures} \label{spinstruct}
%%%%%%%%%%%%%%%%%%%%%%%%%%%%%%%%%%%%%%%%%%%%%%%%%%%%%%%%%%%%

In the NS-R formulation of type II superstring theory, 
the product of functional 
determinants of bosonic, fermionic coordinates and ghosts
for a definite spin structure $s$
is (Right sector)
$$
Z_s = \left( {\theta_s (0) \ov \theta^{'}_1 (0)} \right)^4\ .
$$
The other ingredient which depends on the spin strucure $s$ is 
the fermionic correlator:

$$
\langle \Psi_\mu (z)\Psi_\nu (0)\rangle_s \langle \Psi_\lambda (w)\Psi_\rho
(0)\rangle _s
=g_{\mu\nu}g_{\lambda\rho}{\theta_s (z)\theta_s (w) (\theta^{'}_1 (0))^2
\ov \theta_1 (z)\theta_1 (w)(\theta_s (0))^2}
$$
Summing over spin structures with the GSO phases $\eta_s$, we get
\beq
 \sum_s  \eta_s Z_s \langle \Psi_\mu (z)\Psi_\nu (0)\rangle_s\langle
\Psi_\lambda (w)\Psi_\rho (0)\rangle_s
= g_{\mu\nu}g_{\lambda\rho} 
{\big( \theta_1 ({z+w\ov 2})\big)^2
\big( \theta_1 ({z-w\ov 2}) \big) ^2\ov \theta_1 (z)\theta_1 (w)
\big( \theta^{'}_1 (0) \big)^2}\ .
\eeq
Since $\theta_1 (0)=0$ we obtain a vanishing result 
for $z\rightarrow w$.

Now consider correlators involving $\partial\psi $.
Making first the derivative in $z$ and $w$ 
we get a non-vanishing result for $z\rightarrow w$:
\be
&\lim_{z\to w}&\sum_s  \eta_s Z_s \langle \p\Psi_\mu (z)\Psi_\nu
(0)\rangle_s
\langle \Psi_\lambda (w)\p\Psi_\rho (0)\rangle _s \nonumber \\
&=& g_{\mu\nu}g_{\lambda\rho}
{ \big( \theta_1 (z)\big) ^2  \big(\theta^{'}_1 (0)\big)^2\ov 
\big(\theta_1 (z)\big)^2\big(\theta^{'}_1 (0)\big)^2} =
g_{\mu\nu}g_{\lambda\rho}\ .
\ee 
This is the basic correlator used in appendix \ref{appampl} \ in order
to obtain eq. (\ref{unve}).

%%%%%%%%%%%%%%%%%%%%%%%%%%%%%%%%%%%%%%%%%%%%%%%%%%%%%%%%%%%%
\section{Analytic determination of a decay rate}\label{operatorialdecay}
%%%%%%%%%%%%%%%%%%%%%%%%%%%%%%%%%%%%%%%%%%%%%%%%%%%%%%%%%%%%

We study here the decay of the state $n=k$ in a particular channel
where one of the decay products is massless, by using the operator
formalism.

Let us for short use the following notation for the state $n=k$ 
(in this notation $N=2k-1$):
\beq
|\Phi_{N}\rangle \equiv {1\ov N!} \  \psi^\dagger_1 (b^\dagger )^N 
\tilde\psi^\dagger_2 
({\tilde c}^\dagger )^N|0\rangle\ .
\eeq
We will compute the decay-rate of the process in which
the state $n=k$ decays into a massless particle (``graviton'') 
and another massive state of the 
same kind: 
\beq
|\Phi_{N}\rangle \to |graviton\rangle +|\Phi_{N-l}\rangle
\eeq 
We are interested in the behavior of the decay rate for  
$N\gg 1$ and $l$ finite.

Thus we consider the matrix element
\beq
A=g_s A_RA_L
\eeq
where
\beq
A_R=\langle 0|{b^{N-l}\ov\sqrt{(N-l)!}}\psi_1\ \epsilon\cdot b 
\exp{(-\sqrt{2}p\cdot b)}\ 
{\psi^\dagger_1 {b^\dagger}^N\ov\sqrt{N!}}|0\rangle
\eeq

Here $\epsilon$ is a polarization tensor and 
in complex coordinate notation $\epsilon\cdot b=\epsilon_1 b$
and similarly $p\cdot b=p_1 b$. 

We have written only the part of the graviton vertex which is most relevant
for the $N\to\infty$ limit. Other parts will be suppressed by additional powers of
$p_1\leq p$, the graviton energy being 
$$
p={l\ov 2\sqrt{N}}\ .
$$

The computation of the matrix element gives:
\beq
A_R= \epsilon_1 {(-\sqrt{2}p_1)^{l-1}\ov (l-1)!}\sqrt{N!\ov (N-l)!}\ .
\eeq
and similarly for $A_L$ with $\epsilon_1 ,p_1\to\epsilon_2 ,p_2$.

In the limit $N\to\infty$ at fixed $l$, 
$A_{R,L}\to c(l,\theta )\ \sqrt{N}$
where $c(l,\theta )$ depends on the angle of the graviton momentum with
the (complex) directions $1,2$ respectively.

We thus get the decay rate for the present channel:
\beq
D(N,l)={1\ov\sqrt{N}}g_s^2 \int d\Omega_9 \ 
|A_R\cdot A_L|^2 {p^7\ov \sqrt{N}} \ ,
\eeq
where the last phase space factor comes from
$$
\int {p^8dp\ov p\sqrt{p^2+N-l}}\delta (\sqrt{N}-p-\sqrt{p^2+N-l}~)
= {p^7\ov \sqrt{N}}\ ,
$$
and we have ignored some numerical factors.
The integration over the solid angle gives a suppression factor 
which depends on $l$ but not on $N$.

Finally we get the behavior:
\beq
D(N,l)\cong g_s^2\  c_0(l)\ N^{-5/2}\ ,\ \ \ \ \ N\gg 1\ .
\label{wer}
\eeq
%If we try to investigate the dependence on $l$ we get a distribution 
%which looks
%broadly speaking similar to the one obtained with the precise computation
%reported in Sect. 4, but it is quantitatively substantially different.
We think that other channels are also relevant in the process
$M\to M_1+massless$. In those channels 
%not only the massless particle can be other than
%the graviton, but also 
the massive final state can be different from the class of
``$n=k$'' states considered here.

%This suggests that the decay rate into 
%other relevant channels may also have the behavior
% (\ref{wer}) of the form $\sim g_s^2\  c_0(l)\ M^{-5}$.

%Note that in $d$ dimensions the decay rate would be
%\beq
%D^d(N,l)\cong g_s^2\  c_0(l)\ N^{-(D-5)/2}\ ,\ \ \ \ \ N\gg 1\ .
%\label{ddim}\eeq

\clearpage

%%%%%%%%%%%%%%%%%%%%%%%%%%%%%%%%%%%%%%%%%%%%%%%%%%%%%%%%%%%%%%%%%%%%%%%%%%%%%%%%%%%%
\section{Three-dimensional plot of the logarithm of the decay rate
for $n=0$ and $J_{max}$ cases}\label{S0shape}
%%%%%%%%%%%%%%%%%%%%%%%%%%%%%%%%%%%%%%%%%%%%%%%%%%%%%%%%%%%%%%%%%%%%%%%%%%%%%%%%%%%%

Here we show  graphics representing the shape of the 
logarithm of the decay rate 
(divided by $2M^2$), called $S_0$ in the text, as a function of $M_1^2,M_2^2$.

The figures have been obtained focusing on an interval in the phase
space 
in which 
$0\leq M_{1, \, 2}^2\leq 20$, and limiting the value of $S_0$ 
to a range $[-0.4, 0]$.

We have chosen a perspective so as to have a clear view 
 of the distance between the maxima and the plane $S_0=0$
(cf. also figs. 2 and 3).

%The point of view is set in order to have the line where $M_1=M_2$ 
%frontal with respect to the observer.

 \begin{figure}[hbt]
\label{S0shapesbackn0} 
\vskip -0.5cm %\hskip -1cm
\centerline{\epsfig{figure=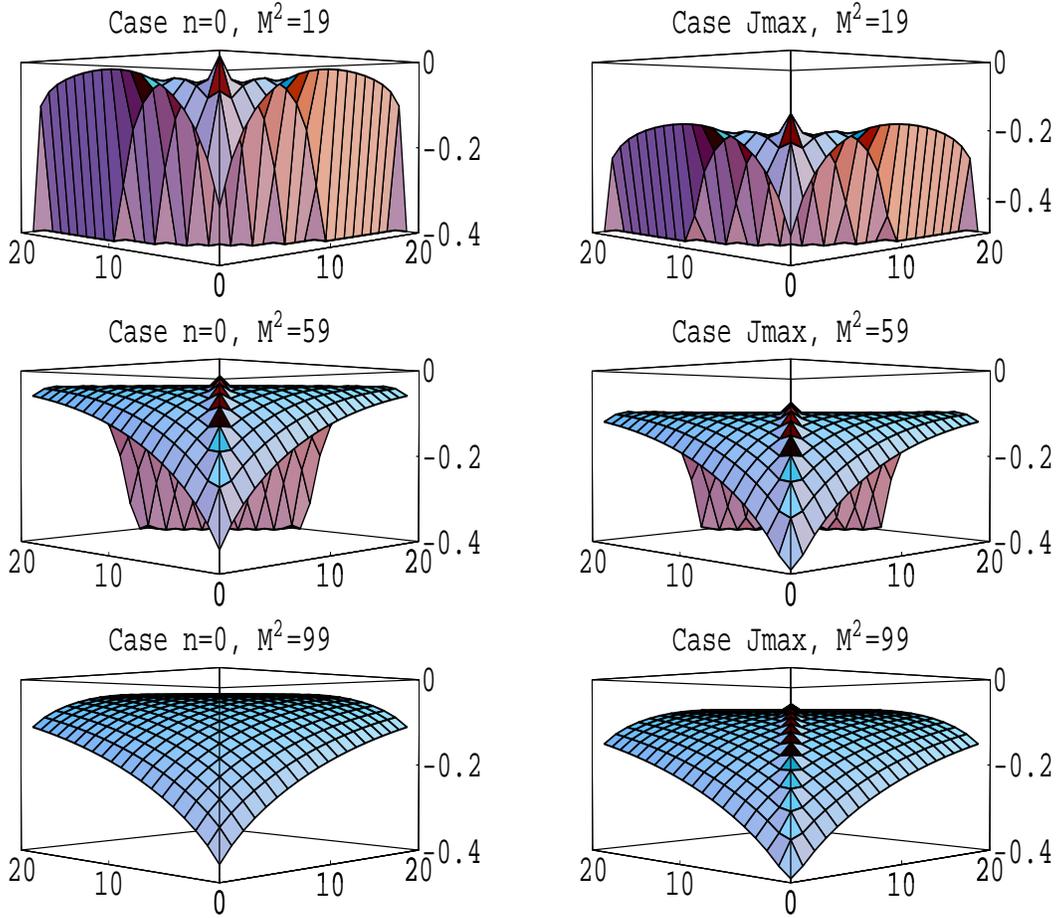,height=13truecm, width=16truecm}}
%\centerline{\includegraphics[width=1in]{splittingt}}
\caption[Three-dimensional plots of $S_0$.]
{\footnotesize
Three-dimensional plot of $S_0$.}
\end{figure}

% \begin{figure}[hbt!]
%\label{S0shapesbackJ} 
%\vskip -0.5cm %\hskip -1cm
%\centerline{\epsfig{figure=S0shapeJ.eps,height=8truecm, width=16truecm}}
%\centerline{\includegraphics[width=1in]{splittingt}}
%\caption[Case $J_{max}$.Three-dimensional plot of S0.]
%{\footnotesize
%Case $J_{max}$.Three-dimensional plot of S0.}
%\end{figure}

\clearpage

%\nocite{*}
%\bibliography{testi}
%\bibliographystyle{plain}

%\vfill\eject

%%%%%%%%%%%%%%%%%%%%%%%%%%%%%%%%%
 \end{document}